\documentclass[prl,twocolumn,showpacs,amsmath,amssymb,superscriptaddress]{revtex4}
\usepackage{graphicx}
\begin{document}
\title{Resonant Spin Polarization and Hall Effects in a Two-Dimensional Electron Gas}
\author{Degang Zhang}
\affiliation{Texas Center for Superconductivity and Department 
of Physics, University of Houston, Houston, TX 77204, USA}
\author{Yao-Ming Mu}
\affiliation{Center for Advanced Materials and Department of Electrical and Computer
Engineering, University of Houston, Houston, TX 77204, USA}
\author{C. S. Ting}
\affiliation{Texas Center for Superconductivity and Department 
of Physics, University of Houston, Houston, TX 77204, USA}

\begin{abstract}

We have studied transport properties in a two-dimensional electron gas with equal Rashba and Dresselhaus spin-orbit interactions under a perpendicular magnetic field. By employing the exact solution for this system, we found resonant charge and spin Hall conductances at a certain magnetic field, where all the nearest-neighboring Landau levels cross. Near this value of magnetic field, there exists a resonant spin polarization, which can also induce resonant charge and spin Hall effects.

\end{abstract}

\pacs{72.20.My, 73.63.Hs, 75.47.-m}

\maketitle

The spin Hall effect is the current focus of theoretical and experimental investigations [1-6]. The Rashba or Dresselhaus spin-orbit interaction plays a fundamental role in generating this effect [7,8,9]. Under a perpendicular magnetic field, a resonant spin Hall conductance is induced in two-dimensional electron gas (2DEG) with pure Rashba coupling due to the crossing of two nearest-neighboring Landau levels while the charge Hall conductance is not influenced [10]. However, the resonant phenomenon could be observed only if this degeneracy happens at the Fermi level. Furthermore, it is difficult to fix experimentally the degenerate point on the Fermi level, which depends on a particular filling factor or density of electrons. In semiconductors, the Rashba and Dresselhaus spin-orbit interactions usually coexist. Because the Rashba and Dresselhaus spin-orbit interactions have different symmetries, the competition between the Rashba and Dresselhaus spin-orbit couplings and the Zeeman energy leads to a novel energy spectrum. When the Rashba and Dresselhaus coupling strengths are equal, the energy spectrum has an equal-distant-like structure and the energy level splitting for the $n$th Landau level is independent of the Landau level number $n$ [11]. Meanwhile, all the energy level crossings between the nearest neighboring Landau levels occur at a fixed magnetic field. This is very different from the case of the pure Rashba coupling in Ref. [10] or unequal Rashba and Dresselhaus coupling strengths [11], where only a single crossing exists at a certain magnetic field. Therefore, 2DEG with equal Rashba and Dresselhaus couplings in a perpendicular magnetic field is more interesting. 
It is well-known that different energy spectrum leads to different physical phenomenon. Here, we investigate spin and charge Hall effects in this special system. Not only  resonant spin Hall conductance is obtained, this resonance behavior  also appears in charge Hall conductance and spin polarization. Because a degeneracy always exists at the Fermi level and the spin polarization is a detectable quantity, these resonant phenomena could be easily observed  and may have potential application for spintronics.

The Hamiltonian for a single electron with spin-$\frac{1}{2}$ in a plane under a
perpendicular magnetic field is given by
$$ H=\frac{1}{2m^*}{\bf \Pi}^2-\frac{1}{2}g_s\mu_BB\sigma_z
+\frac{\alpha}{\hbar}(\sigma_y\Pi_x-\sigma_x\Pi_y)$$
 $$  +\frac{\beta}{\hbar}(\sigma_x\Pi_x-\sigma_y\Pi_y),\eqno{(1)}$$
where ${\bf \Pi}={\bf p}+\frac{e}{c}{\bf A}$, $\sigma_i(i=x, y, z)$ are the Pauli matrices 
for electron spin, $g_s$ is the Lande g-factor, 
$\mu_B$ is the Bohr magneton, $\alpha$ and $\beta$ represent the Rashba and the Dresselhaus spin-orbit couplings,
respectively. Here we choose the Landau guage ${\bf A}=yB\hat{{\bf x}}$. Note that $p_x=k$ is a good quantum number due to $[p_x, H]=0$. 

The Rashba and Dresselhaus spin-orbit interactions originate from the lack of structure and bulk inversion symmetries, respectively [7,8]. The Rashba coupling $\alpha$ can be adjusted by a gate voltage perpendicular to the 2DEG plane [12,13]. Therefore, in experiments, an arbitrary ratio of two kinds of spin-orbit couplings can be obtained in different samples by changing the gate voltage. The relative strength of the Rashba and Dresselhaus couplings can be extracted from the photocurrent measurements [14,15]. Usually the coefficients $\alpha$ and $\beta$ have the same order of magnitude in quantum wells such as GaAs while in narrow gap compounds such as InAs, $\alpha$ dominates [12,13].

When $B=0$, the Hamiltonian (1) has been studied by a lot of authors [16-19]. In Refs. [16,17], the spin Hall effect was investigated by using the Heisenberg equation of motion and the Kubo formalism approaches, respectively. The out of plane spin current only depends on the sign of $\beta^2-\alpha^2$ for $\alpha\not=\beta$ and vanishes at $\alpha=\beta$. We note that when $\alpha=\beta$, the Hamiltonian (1) has some special symmetries and its spin state of the wavefunction is independent of the wave vector. This property was proposed to design a non-ballistic spin-field-effect transistor [18]. In addition, due to an exact SU(2) spin rotation symmetry, a Persistent Spin Helix, whose lifetime is infinite, was discovered at this point [19].  

When $B\not=0$, the Hamiltonian (1) with a pure Rashba coupling (i.e. $\beta=0$) was solved by Rashba over fourty years ago [7]. The Landau levels of 2DEG with a pure Dresselhaus coupling (i.e. $\alpha=0$) can be obtained by the unitary transformation $\sigma_x\rightarrow\sigma_y$, $\sigma_y\rightarrow\sigma_x$ and $\sigma_z\rightarrow -\sigma_z$, which maps a 2DEG with Rashba coupling $\alpha$, Dresselhaus coupling $\beta$, and Lande g-factor $g_s$ to a 2DEG with Rashba coupling $\beta$, Dresselhaus coupling $\alpha$, and Lande g-factor $-g_s$ [20,21]. 
When the Rashba and Dresselhaus couplings coexist, the Hamiltonian (1) becomes very complicated and its solution is not obtained  for many years. Very recently, the problem has been solved by using unitary transformations and introducing two bosonic annihilation operators $b_{k\sigma}=\frac{1}{\sqrt{2}l_c}[y+\frac{c}{eB}(k+ip_y)+\sqrt{|a_R a_D|}u_\sigma]$ and the corresponding creation operators $b_{k\sigma}^\dagger=(b_{k\sigma})^\dagger$, with the cyclotron radius $l_c=\sqrt{\frac{\hbar c}{eB}}$, $a_R=\frac{\alpha m^*l_c}{\hbar^2}$, $a_D=\frac{\beta m^*l_c}{\hbar^2}$, $u_\sigma=\sigma[1-i {\rm sgn}(\alpha\beta)]$, and the orbital index $\sigma=\pm 1$ [11]. Here we point out that two constants $\sqrt{|a_R a_D|}u_\sigma$ in the operators $b_{k\sigma}$ play an important role in solving the Hamiltonian (1). Obviously, different from the pure Rashba or Dresselhaus coupling case, the orbital space of electron is divided into two independent infinite-dimensional subspaces described by the occupied number representations $\Gamma_\sigma$ associated with $b_{k\sigma}$ and $b^\dagger_{k\sigma}$. Then the Hamiltonian (1) is rewritten as $H=H_{-1}\oplus H_1$, where $H_\sigma$ is the sub-Hamiltonian in $\Gamma_\sigma$. The exact solution for the Hamiltonian (1) can be expressed as an infinite series in terms of the free Landau levels in each $\Gamma_\sigma$ and the physical parameters. The energy level splitting is dependent on the Landau level number when $|a_R|\not=|a_D|$ while the energy spectrum has an equal-distant-like structure if $|a_R|=|a_D|$. 
When the spin-orbit couplings $\alpha$ and $\beta$ are small, the exact results are consistent with those obtained by the perturbation theory and truncation approximation [21,22]. In this work, we focus on the magnetic and transport properties of the Hamiltonian (1) at $|a_R|=|a_D|$ (i.e. $|\alpha|=|\beta|$), which can be realized by the photocurrent experiment [14,15]. 

When $|a_D|=|a_R|=a$, the eigenvalue for the $n$th Landau level with the spin index $s$ and the orbital index $\sigma$ is given by [11]
$$E_{ns\sigma}=\hbar\omega(n+\frac{1}{2}+2a^2+\frac{1}{2}s\sqrt{g^2+64a^4}).\eqno{(2)}$$
where $s=\pm 1$, the cyclotron frequency $\omega=\frac{eB}{m^*c}$, and $g=\frac{g_sm^*}{2m_e}$ with $m_e$ the free electron mass. Note that $E_{ns\sigma}$ is independent of the orbital index $\sigma$. In other words, there is the same energy spectrum in each $\Gamma_\sigma$. Obviously, the energy splitting induced by the Zeeman energy and spin-orbit coupling is independent of the Landau level index $n$, i.e. $\Delta E=E_{ns=1}-E_{ns=-1}=\hbar\omega\sqrt{g^2+64a^4}$. In Fig. 1, we present several low-lying Landau levels changing with the parameter $a$ when $g=0.1$. Obviously, all the nearest neighboring Landau levels cross at $a^*=\frac{1}{2\sqrt{2}}(1-g^2)^{\frac{1}{4}}$.

\begin{figure}
\rotatebox[origin=c]{0}{\includegraphics[angle=0, 
           height=2.0in]{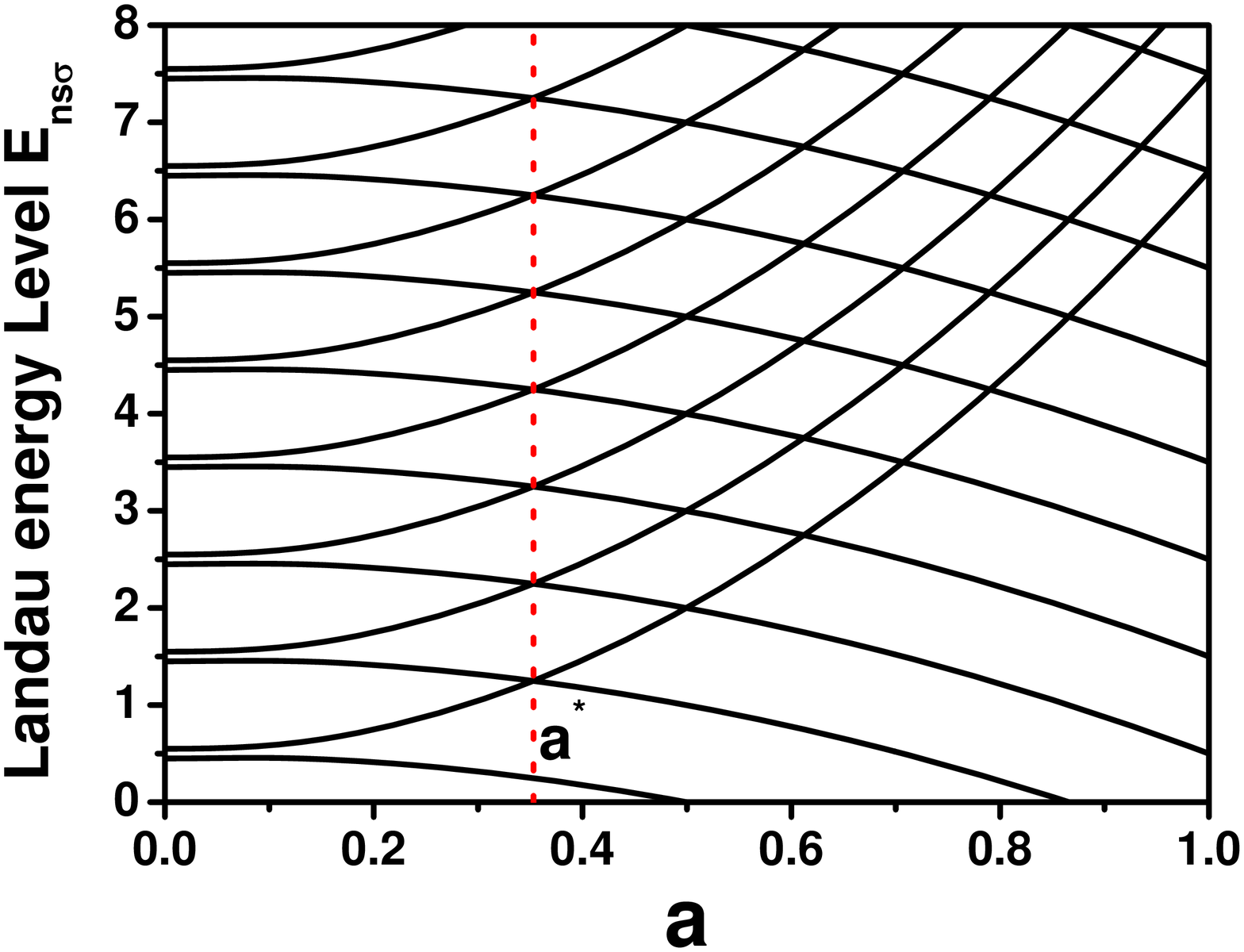}}                   
\caption {(Color online). Landau energy levels of an electron as functions of $a$ with $g=0.1$ in the unit of $\hbar\omega$. All the crossings of the nearest neighboring Landau levels locate on the dot line at $a^*$, which give rise to resonant spin and charge Hall conductances. Near the point (slightly larger than $a^*$), there is a resonant spin polarization, which can also induce resonant spin and charge Hall effects.}
\end{figure}

The eigenstate corresponding to the eigenvalue (2) reads [11]
$$|n,k,s,\sigma>=\frac{1}{{\cal A}_{ns\sigma}}\sum_{m=0}^{+\infty}
\alpha_{ms}^{n\sigma}
\left (
\begin{array}{c}
1\\
\lambda_{nms}T_\sigma
\end{array}\right )\phi_{mk\sigma},$$
$$\lambda_{nms}=\frac{n-m+4a^2+\frac{1}{2}g+\frac{1}{2}s\sqrt{g^2+64a^4}}
{n-m+4a^2-\frac{1}{2}g+\frac{1}{2}s\sqrt{g^2+64a^4}},$$
$$T_\sigma=\frac{\sqrt{2}}{2}\sigma({\rm sgn}\beta+i{\rm sgn}\alpha),$$
$$|{\cal A}_{ns\sigma}|^2=\sum_{m=0}^\infty(1
+\lambda_{nms}^{2})|\alpha_{ms}^{n\sigma}|^2,\eqno{(3)}$$ 
where $\phi_{mk\sigma}$ is the eigenstate of the $m$th Landau level in $\Gamma_\sigma$, i.e. $b_{k\sigma}^\dagger \phi_{mk\sigma}=\sqrt{m+1}\phi_{m+1k\sigma}$, $b_{k\sigma} \phi_{mk\sigma}=\sqrt{m}\phi_{m-1k\sigma}$, $<\phi_{m^\prime k\sigma^\prime}|\phi_{mk\sigma}>=\delta_{mm^\prime}\delta_{\sigma\sigma^\prime}$, ${\it lim}_{m \rightarrow + \infty}  \alpha_{ms}^{n\sigma}=0$, and the coefficient $\alpha_{ms}^{n\sigma}$ satisfies the recursion relation
$$\sqrt{m}au_\sigma(1+\lambda_{nm-1s})\alpha_{m-1s}^{n\sigma}
+\sqrt{m+1}au^*_\sigma(1+\lambda_{nm+1s})\alpha_{m+1s}^{n\sigma}$$
$$-(n-m+\frac{1}{2}g+\frac{1}{2}s\sqrt{g^2+64a^4}-4a^2\lambda_{nms})\alpha_{ms}^{n\sigma}=0.\eqno{(4)}$$
Obviously, the expectation value of any physical quantity in the $n$th Landau level is completely determined by $\lambda_{nms}$ and $\alpha_{ms}^{n\sigma}$. 
When $g=0$, from Eq. (6), we have the exact solutions: (i) $\lambda_{nns=-1}=-1$, $\alpha_{ns=-1}^{n\sigma}=1$ and $\alpha_{ms=-1}^{n\sigma}=0$ for $m\not =n$; (ii) $\lambda_{nms=1}=1$ and $\alpha_{ns=1}^{m\sigma}$ satisfy the relation (4) with the replacement of parameters. It is easy to prove that when $g=0$, there is no spin polarization in each Landau level, i.e. $<n,k,s,\sigma|\frac{\hbar}{2}\sigma_z|n,k,s,\sigma>=0$. So it is not expected that there exists an out of plane spin current. In this case, the orbital and spin wave functions of electrons are separated, and the spin part $\frac{\sqrt{2}}{2}(1~sT_\sigma)^T$ is independent of the wave vector $k$, which is similar to that observed in the absence of a magnetic field [18]. In the following, we calculate spin polarization, spin and charge Hall conductances in the Hamiltonian (1) with $g\not=0$ at high and low magnetic fields by using the exact energy spectrum described by Eqs. (2)-(4).

Using the eigenstates (3) and the recursion relation (4), the expectation value of spin polarization per electron is
$$S_z=\frac{1}{2\nu}\sum_{ns\sigma}<n,k,s,\sigma|\frac{\hbar}{2}\sigma_z|n,k,s,\sigma>f(E_{ns\sigma})$$
$$=\frac{\hbar}{4\nu}\sum_{nms\sigma}\frac{1}{|{\cal A}_{ns\sigma}|^2}(1-\lambda_{nms}^2)|\alpha_{ms}^{n\sigma}|^2f(E_{ns\sigma}),
\eqno{(5)}$$
where $\nu$ is the filling factor and $f(x)$ is the Fermi distribution function. Note that two in the denominator of the first expression of $S_z$ comes from two orbital subspaces. 
In Figs. 2(a) and 2(d), we show the zero temperature spin polarization curves at high and low magnetic fields, respectively. In our calculations, we have chosen the parameters $\alpha=\beta=0.9\times 10^{-11}$ eVm, $n_e=1.9\times 10^{16}/m^2$, $g_s=4$, and $m^*=0.05m_e$, taken for In$_{0.53}$Ga$_{0.47}$As/In$_{0.52}$Al$_{0.48}$As quantum well [12,15]. 
The purposes of this parameter choice are to compare with the results in 2DEG with pure Rashba coupling in Refs. [10,21] and to understand the role of the competition between the spin-orbit interactions and the Zeeman energy. However, we verified that varying these parameters in a reasonable range does not qualitatively change our results. Because of the particular symmetry in this system, the spin polarization has many special features, compared to the case of the pure Rashba or Dresselhaus coupling. At high magnetic fields, the spin polarization almost vanishes between 5T $\sim$ 20T. This is the result of the competition between the Zeeman energy and the spin-orbit couplings. Below 5T, the spin polarization oscillates with the magnetic field due to the alternate filling of the Landau levels. At low magnetic fields, we can see in Fig. 2(d) that $S_z$ has a finite jump at the energy level crossing $a^*$ ($B\approx\frac{1}{5.44}$T), similar to the case of the pure Rashba coupling [10]. More interestingly, a resonant spin polarization is discovered near the degenerate point, i.e. $B\approx\frac{1}{5.46}$T, which is one of the main results in this paper. This resonant phenomenon is produced by the property of the quantity $\lambda_{nms}$. We shall see below that the resonant spin polarization induces resonant spin and charge Hall effects, which is never reported before.

In order to investigate the spin and charge transport properties in 2DEG, we apply a tiny electric field $E$ along the $y$ axis. So the potential energy $H^\prime=eEy$, treated as a perturbation term, must be added into the Hamiltonian (1). The operator $y$ can be expressed in terms of the bosonic operators $b_\sigma$ and $b_\sigma^\dagger$ in each subspace $\Gamma_\sigma$, i.e. $y=\frac{\sqrt{2}}{2}l_c(b_{k\sigma}^\dagger+b_{k\sigma})-\frac{ck}{eB}-\sigma a$. The charge current operator of a single electron in $\Gamma_\sigma$ reads $ j_{c\sigma}=-ev_{x\sigma}$
while the corresponding out of plane spin current operator in $\Gamma_\sigma$ is 
$j_{sz\sigma}=\frac{\hbar}{4}(\sigma_z v_{x\sigma}+v_{x\sigma}\sigma_z)$.
Here, the electron velocity in $\Gamma$ along $x$ axis $v_{x\sigma}=\frac{1}{i\hbar}[x,H_\sigma+H^\prime_\sigma]=\frac{\hbar}
{\sqrt{2}m^*l_c}[b_{k\sigma}^\dagger+b_{k\sigma}+\sqrt{2}a(\sigma_y {\rm sgn}\alpha+\sigma_x {\rm sgn}\beta-\frac{\sigma}{l_c})]$.
Up to the first order in the electric field $E$ in the expectation value of the spin or charge current operator, 
then spin or charge Hall conductance, i.e. the coefficient of the linear term,  can be expressed as [10,22]

\begin{widetext}
$$G_{sz,c}=\frac{1}{E}\sum_{nn^{\prime}kss^{\prime}\sigma}[\frac{<n^\prime,k,s^\prime,\sigma|H^\prime_\sigma|n,k,s,\sigma><n,k,s,\sigma|j_{sz\sigma,c\sigma}|n^\prime,k,s^\prime,\sigma>}
  {E_{ns\sigma}-E_{n^\prime s^\prime \sigma}}f(E_{ns\sigma})+{\rm h.c.}],
\eqno{(6)}$$
\end{widetext}
where the matrix elements $<n^\prime,k,s^\prime,\sigma|H^\prime_\sigma|n,k,s,\sigma>$, $<n^\prime,k,s^\prime,\sigma|j_{sz\sigma}|n,k,s,\sigma>$, and $<n^\prime,k,s^\prime,\sigma|j_{c\sigma}|n,k,s,\sigma>$ can be easily obtained by using the eigenstate (3). Obviously, $G_{sz}$ ang $G_{c}$ are highly nonlinear functions in terms of the parameters $a$ and $g$, which reveal abundant transport characteristics in this special system. We note that when $a\rightarrow 0$, Eq. (6) is consistent with that in 2DEG with pure Rashba (Dresselhaus) spin-orbit interaction in the limit of vanishing coupling strength [10,21].  
 
\begin{figure}
\rotatebox[origin=c]{0}{\includegraphics[angle=0, 
           height=1.18in]{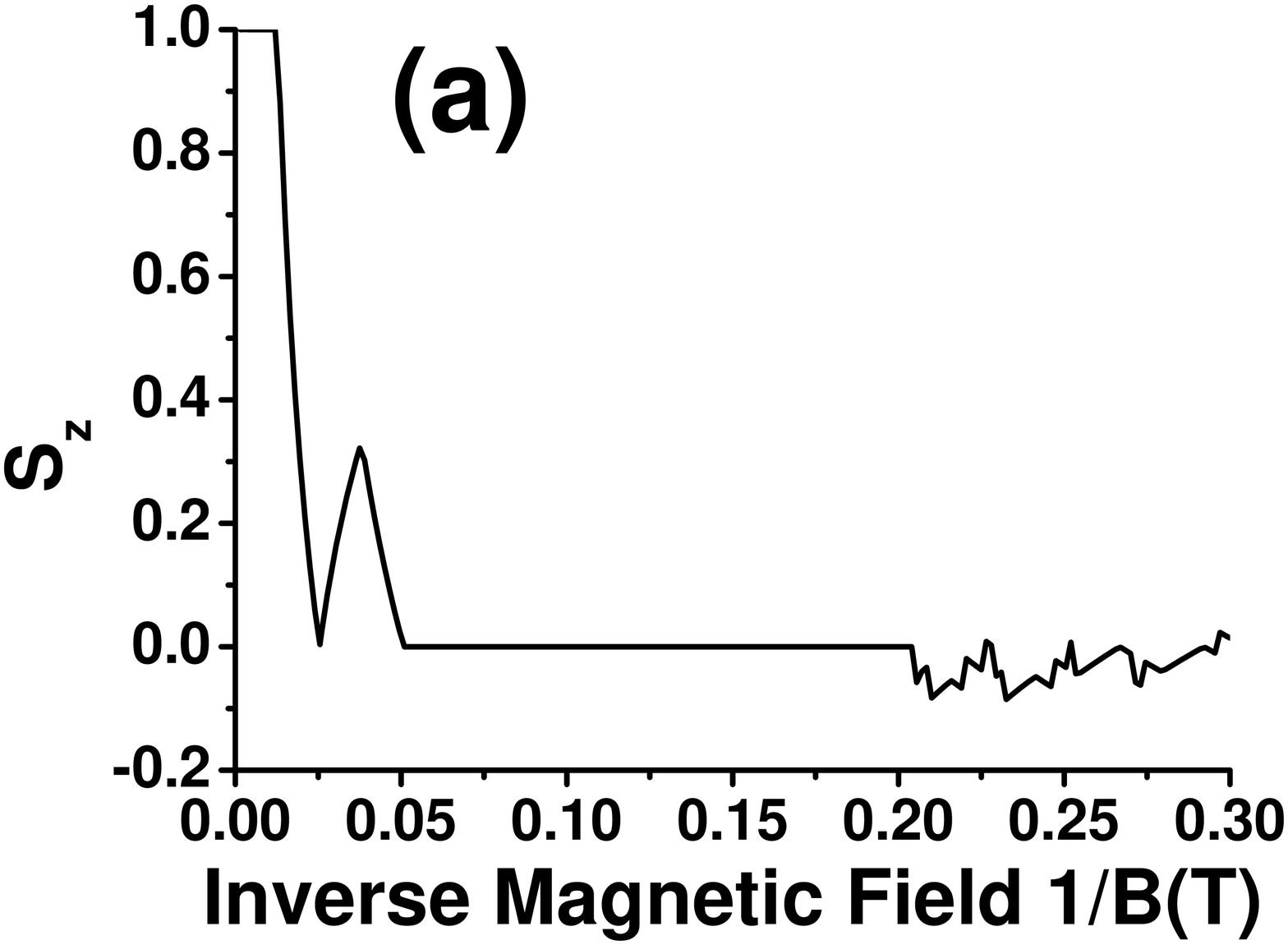}}
\rotatebox[origin=c]{0}{\includegraphics[angle=0, 
           height=1.18in]{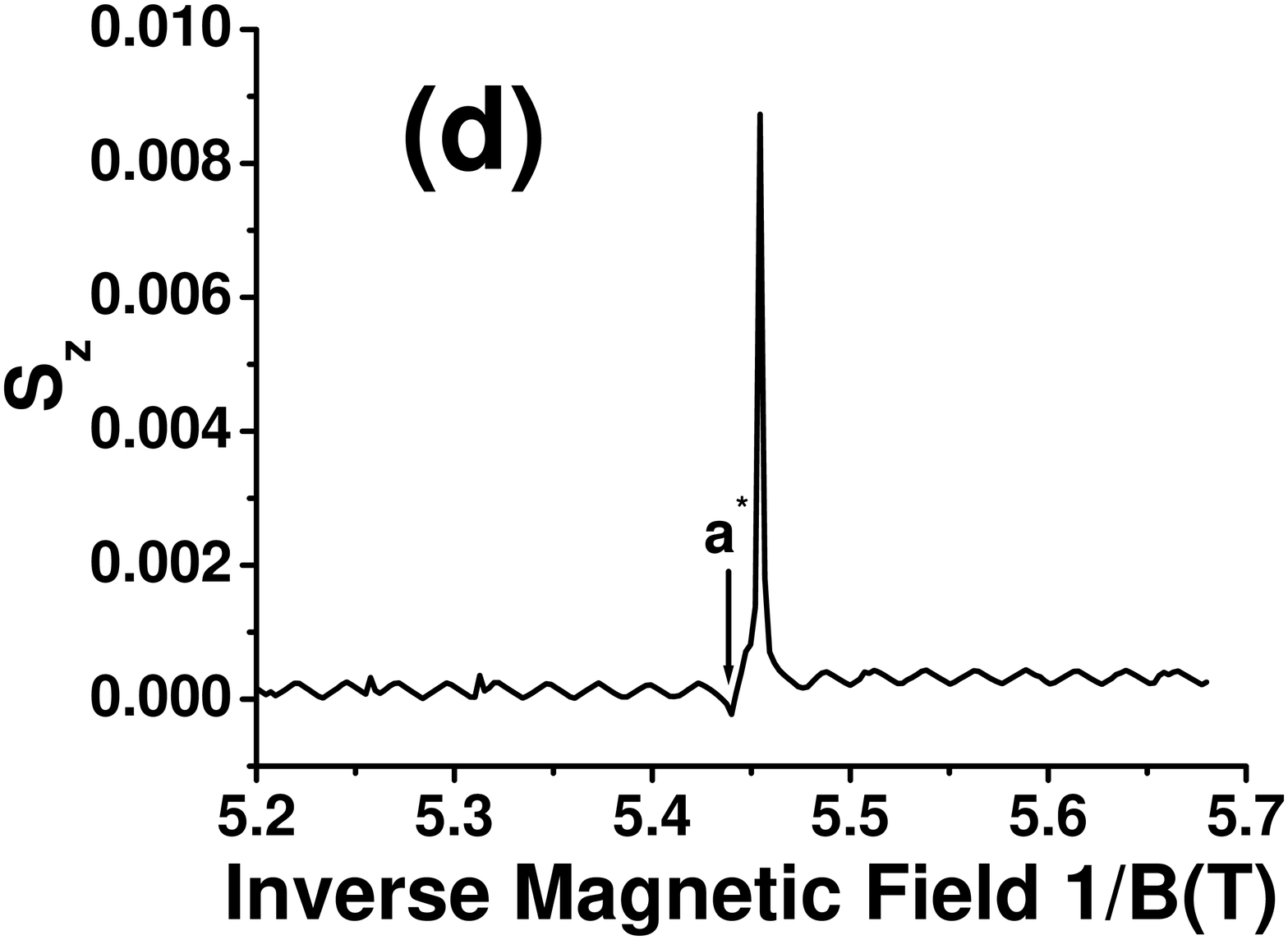}}
\rotatebox[origin=c]{0}{\includegraphics[angle=0, 
           height=1.18in]{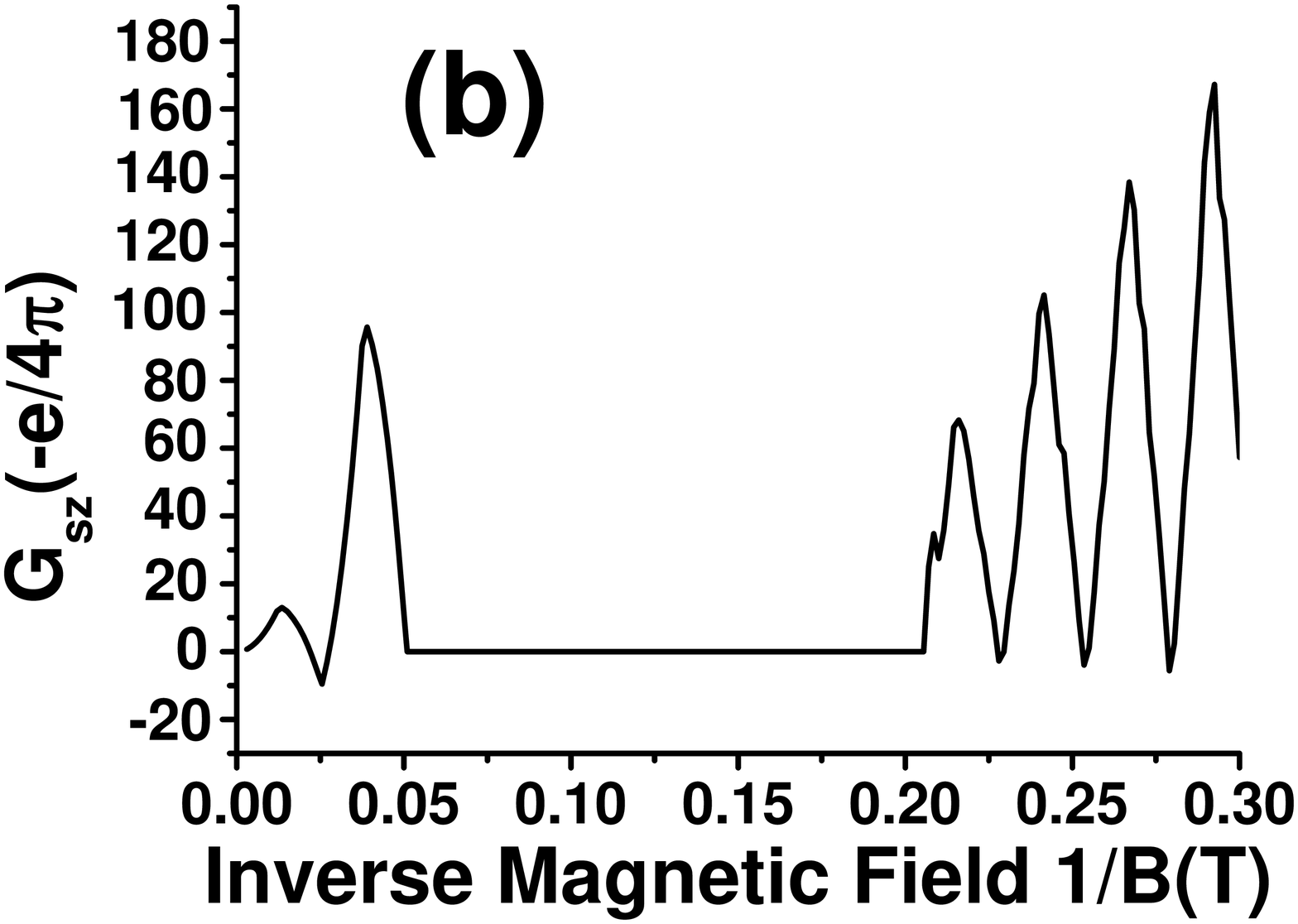}}
\rotatebox[origin=c]{0}{\includegraphics[angle=0, 
           height=1.18in]{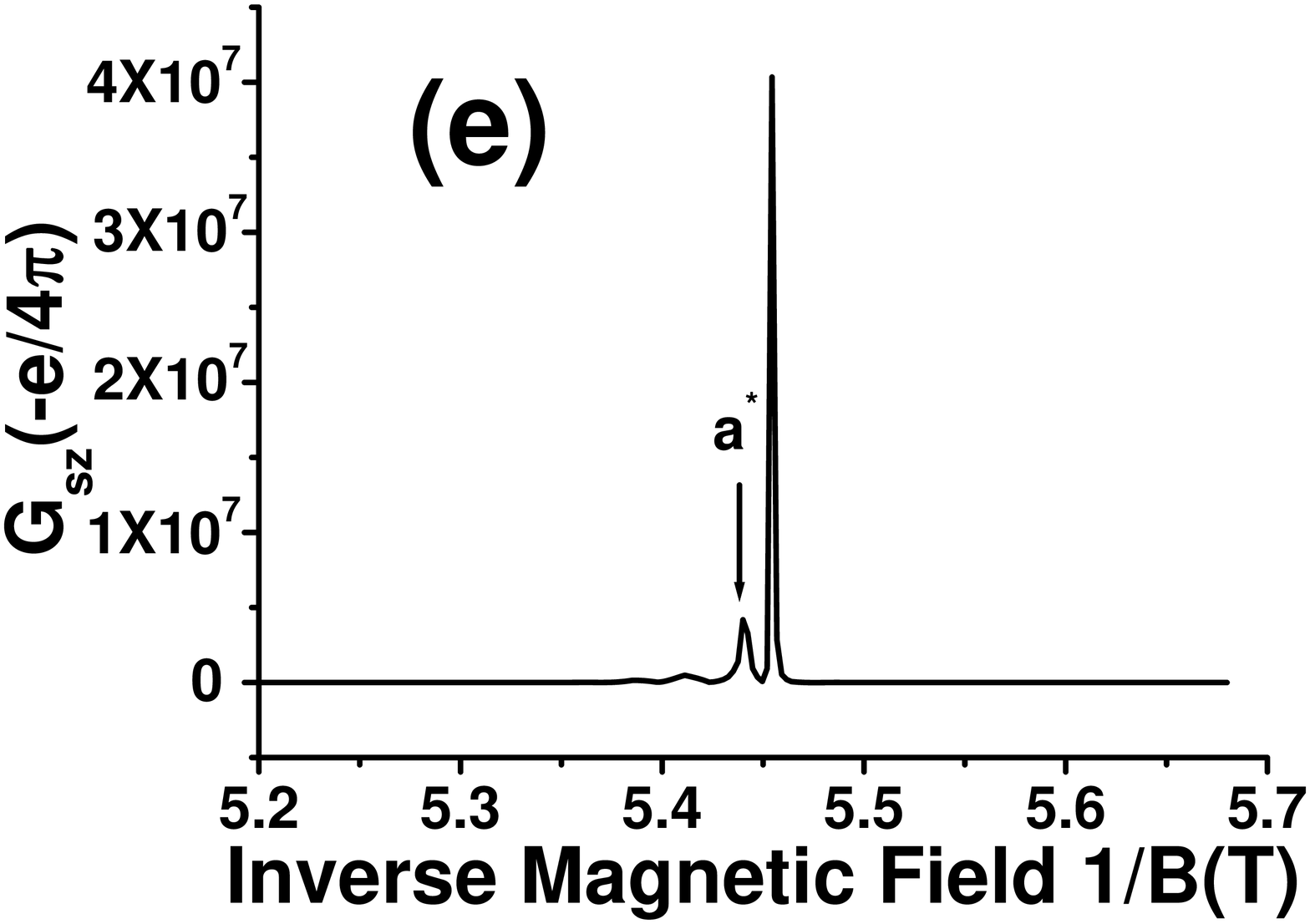}}
\rotatebox[origin=c]{0}{\includegraphics[angle=0, 
           height=1.18in]{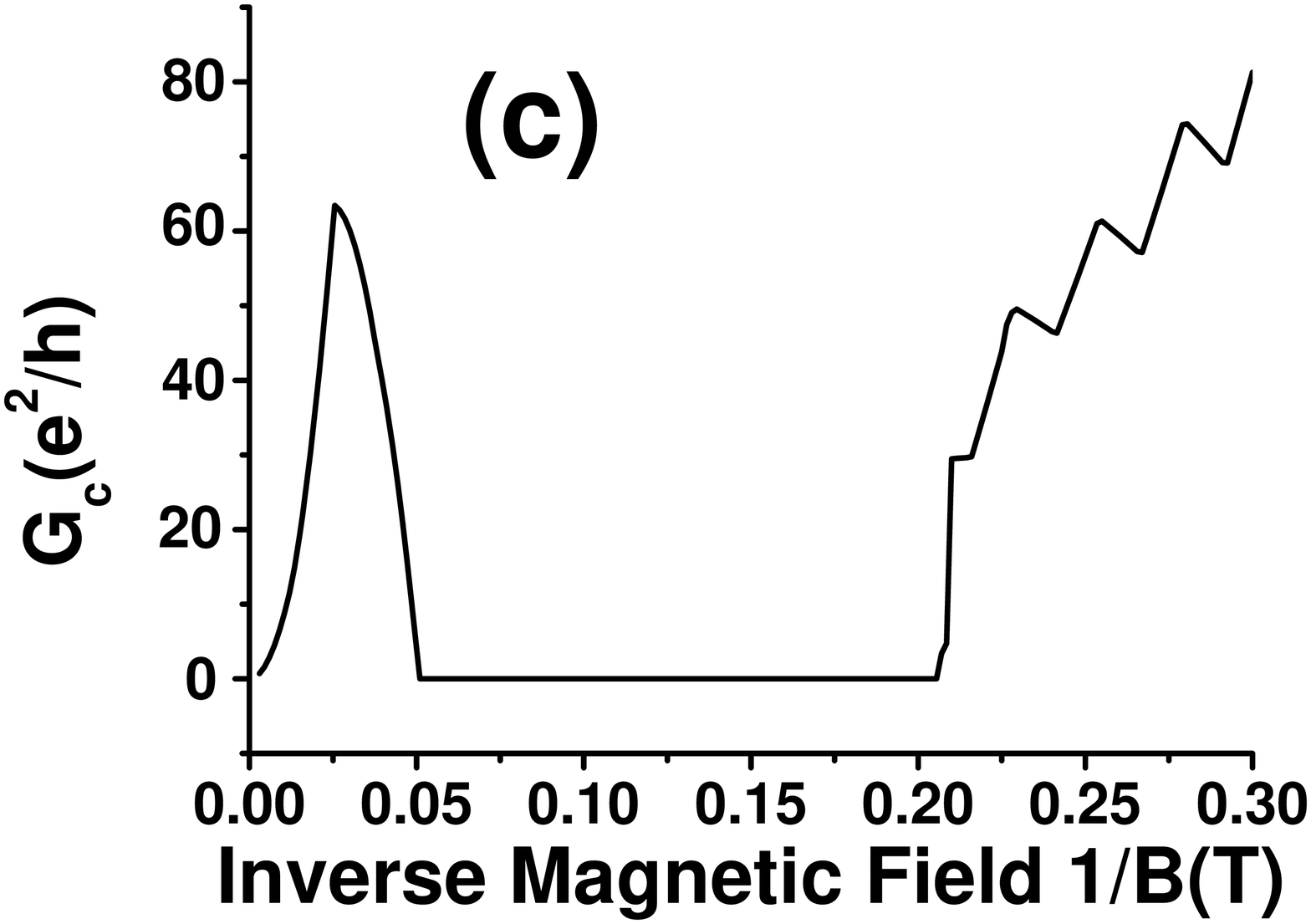}} 
\rotatebox[origin=c]{0}{\includegraphics[angle=0, 
           height=1.18in]{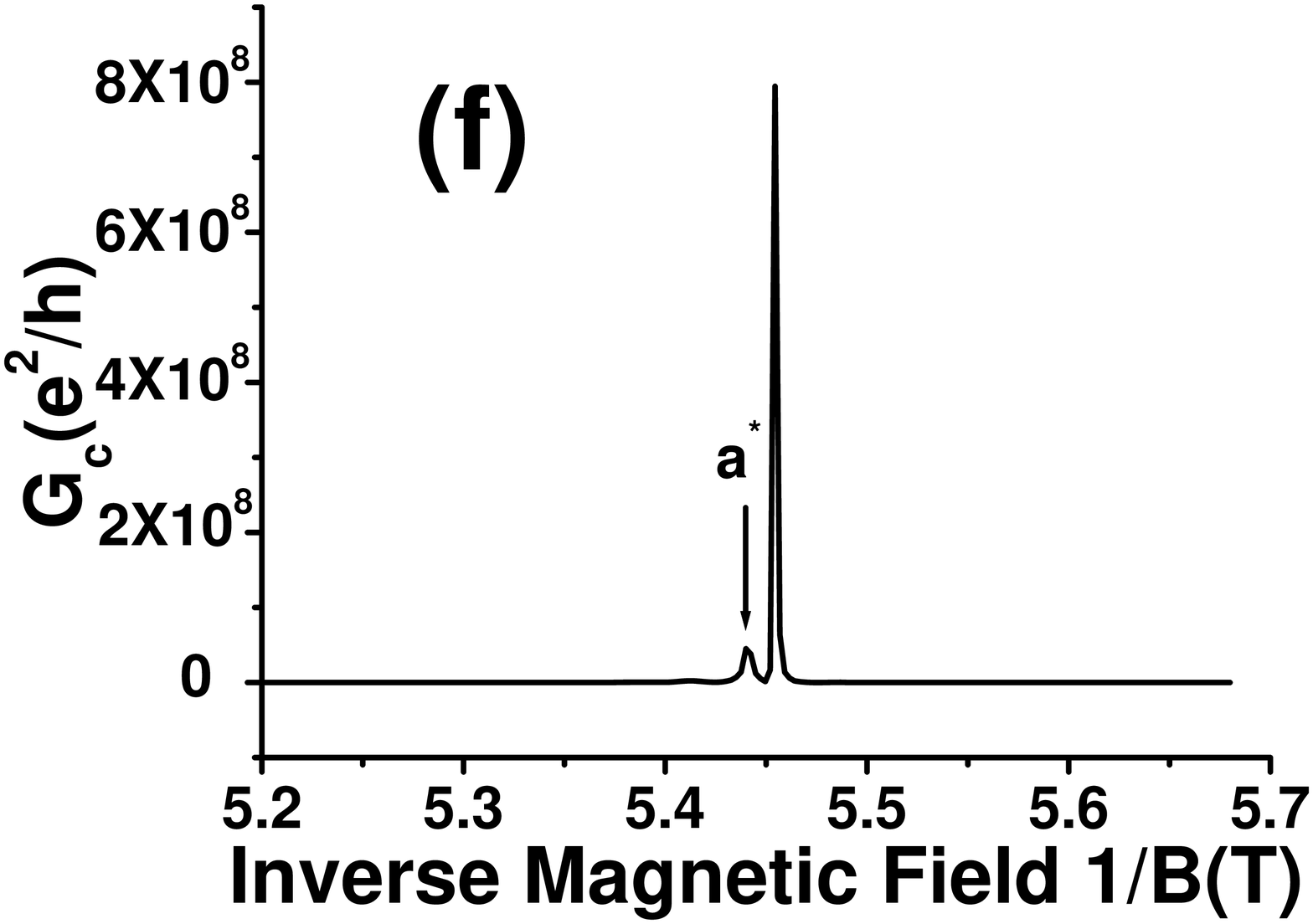}}              
\caption{$S_Z$ (unit: $\hbar/2$), $G_{sz}$ and $G_c$ as a function of $1/B$ at zero temperature. The sample parameters used here are $\alpha=\beta=0.9\times 10^{-11}$ eVm, $n_e=1.9\times 10^{16}/m^2$, $g_s=4$, and $m^*=0.05m_e$.}           
\end{figure}

According to the formula (6), we plot the curves of $G_{sz}$ and $G_{c}$ at zero temperature as a function of $\frac{1}{B}$ in Figs. 2(b), 2(e), 2(c) and 2(f). Similar to the curve of $S_z$, $G_{sz}$ and $G_{c}$ almost vanish in the magnetic field interval [5T, 20T). This feature does not appear in the spin and charge Hall conductances in 2DEG with pure Rashba or Dresselhaus coupling [10,21].
Below 5T, $G_{sz}$ possesses an  amplitude varying saw tooth structure with a period $\sim 0.024T^{-1}$ while  $G_{c}$ has a step-like form with the same period. Contrary to the case of the pure Rashba coupling, the charge Hall conductance $G_c$ strongly depends on the spin-orbit coupling strength. Obviously, in low magnetic field region shown in Figs. 2(e) and 2(f), both $G_{sz}$ and $G_{c}$ have two resonant peaks at and near $a^*$, which are induced by the energy level crossings and the resonant spin polarization, respectively. The resonant peak of $G_{sz}$ and $G_{c}$ induced by the resonant spin polarization is much higher than that produced by the energy level crossings. Here we point out that very different from the case of the pure Rashba coupling, the resonant spin Hall effect at $a^*$ is easily observed by experiments because an energy level crossing always exists at the Fermi level. When $B\rightarrow 0$, the spin Hall conductance vanishes, which recovers the result in the absence of a magnetic field [16,17].

In summary, we have studied the spin polarization, spin and charge Hall conductances in 2DEG in the presence of equal Rashba and Dresselhaus spin-orbit interactions under a perpendicular magnetic field by employing the exact solution for the Hamiltonian (1). Besides the resonant phenomena at the energy level crossing, we have discovered a  resonance spin polarization near the degenerate point. The resonant spin polarization can lead simultaneously to both resonant spin and charge Hall conductances. Furthermore, the resonant charge Hall effect is not found in the other semiconductor systems up to now. We emphasize that the results obtained here depend only on the magnitudes of the spin-orbit couplings $\alpha$ and $\beta$ and are independent of their signs. It is expected that these resonant phenomena could be verified by experiments and be used to design devices in spintronics.

This work was supported by the Texas Center for Superconductivity at the University of Houston and by the Robert A. Welch Foundation under grant No. E-1146.

\end{document}